\magnification=1200
\def\qed{\unskip\kern 6pt\penalty 500\raise -2pt\hbox
{\vrule\vbox to 10pt{\hrule width 4pt\vfill\hrule}\vrule}}
\centerline{CHARACTERIZATION OF LEE-YANG POLYNOMIALS.}
\bigskip\bigskip
\centerline{by David Ruelle\footnote{*}{Math. Dept., Rutgers University, and 
IHES, 91440 Bures sur Yvette, France. email: ruelle@ihes.fr}.}
\bigskip\bigskip\bigskip\bigskip\noindent
	{\leftskip=2cm\rightskip=2cm\sl Abstract.  The Lee-Yang circle theorem describes complex polynomials of degree $n$ in $z$ with all their zeros on the unit circle $|z|=1$.  These polynomials are obtained by taking $z_1=\ldots=z_n=z$ in certain multiaffine polynomials $\Psi(z_1,\ldots,z_n)$ which we call Lee-Yang polynomials (they do not vanish when $|z_1|,\ldots,|z_n|<1$ or $|z_1|,\ldots,|z_n|>1$).  We characterize the Lee-Yang polynomials $\Psi$ in $n+1$ variables in terms of polynomials $\Phi$ in $n$ variables (those such that $\Phi(z_1,\ldots,z_n)\ne0$ when $|z_1|,\ldots,|z_n|<1$).  This characterization gives us a good understanding of Lee-Yang polynomials and allows us to exhibit some new examples.  In the physical situation where the $\Psi$ are temperature dependent partition functions, we find that those $\Psi$ which are Lee-Yang polynomials for all temperatures are precisely the polynomials with pair interactions originally considered by Lee and Yang.\par}
\bigskip\bigskip\noindent
{\tt The present paper was submitted for publication in March 2008, but I did\break not post it on the internet.  I do this now, since in the mean time other\break papers continue to appear.  Let me mention in particular "The Lee-Yang and\break Polya-Schur Programs" by J. Borcea and P. Br\"and\'en\break (archiv.org/abs/0809.0401 and 0809.3087).  This work, while disjoint in\break content from the present paper, has in particular the interest of giving\break an extensive bibliography of the mathematics related to the Lee-Yang\break Circle Theorem.}

\vfill\eject
\noindent
{\bf 0. Introduction}
\medskip
	The Lee-Yang circle theorem [3] states that certain polynomials $P$ of degree $n$ in one complex variable $z$ have all their zeros on the unit circle $|z|=1$.  The polynomials $P$ are obtained by taking $z_1=\ldots=z_n=z$ in $\Psi(z_1,\ldots,z_n)$ when $\Psi$ is a {\it Lee-Yang polynomial}, {\it i.e.}, belongs to a certain class $LY_n$ of complex polynomials separately of degree $1$ in $n$ variables (multiaffine polynomials).  Specifically, $LY_n$ consists of those $\Psi$ such that $\Psi(z_1,\ldots,z_n)\ne0$ when $|z_1|,\ldots,|z_n|<1$ and when $|z_1|,\ldots,|z_n|>1$ (including $|z_i|=\infty$ in a sense to be made precise later).  Our current understanding of Lee-Yang polynomials is based on the concept of {\it Asano contraction} [1].  We shall define an {\it inner radius} associated with a multiaffine polynomial $\Phi$, and see that it behaves supermultiplicatively with respect to Asano contraction (Proposition 2).  Using the properties of the inner radius, we shall characterize the $\Psi\in LY_{n+1}$ ($n+1$ variables) in terms of polynomials $\Phi$ in $n$ variables such that $\Phi(z_1,\ldots,z_n)\ne0$ when $|z_1|,\ldots,|z_n|<1$ (Theorem 3).  This characterization will give us a good understanding of $LY_{n+1}$ (Proposition 5), and allow us to exhibit elements of $LY_{n+1}$ outside of the (pair interaction) class originally considered by Lee and Yang (see in particular Example 7(d)).  The original Lee-Yang class is obtained for ''temperature dependent polynomials'' by imposing the Lee-Yang condition at all temperatures (Theorem 9).  For $n>1$, this class is a lower dimensional set in $LY_{n+1}$.  In Section 10 we discuss the situation where multiaffine polynomials are replaced by polynomials of higher degree in each variable.  We conclude by briefly presenting a physical interpretation of the results obtained (Section 11). 
\medskip\noindent
{\bf 1. Definitions.}
\medskip
	Let ${\cal A}_n\subset{\bf C}[z_1,\ldots,z_n]$ consist of the multiaffine polynomials, {\it i.e.}, those which are separately of degree $1$ in each variable $z_1,\ldots,z_n$, with $n\ge1$.  A polynomial $\Phi\in{\cal A}_n$ is thus of the form
$$	\Phi(z_1,\ldots,z_n)=\sum_{X\subset[n]}E_Xz^X      $$
where we have written $[n]=\{1,\ldots,n\}$ and $z^X=\prod_{x\in X}z_x$.
\medskip
	We associate with $\Phi$ the polynomial $\Phi^\dagger$ such that
$$	\Phi^\dagger(z_1,\ldots,z_n)=\sum_{X\subset[n]}E_{[n]\backslash X}^*z^X     $$
and note that
$$	\Phi^\dagger(z_1,\ldots,z_n)=z_1\cdots z_n\Phi(z_1^{*-1},\ldots,z_n^{*-1})^*  $$
so that if $|\alpha_1|=\ldots=|\alpha_n|=1$ we have $|\Phi^\dagger(\alpha_1,\ldots,\alpha_n)|=|\Phi(\alpha_1,\ldots,\alpha_n)|$.
\medskip
	If $\Phi_1,\Phi_2\in{\cal A}_n$, with
$$	\Phi_1(z_1,\ldots,z_n)=\sum_{X\subset[n]}E_X^1z^X\qquad,\qquad
	\Phi_2(z_1,\ldots,z_n)=\sum_{X\subset[n]}E_X^2z^X      $$
we define a {\it convolution} product $\Phi_1*\Phi_2\in{\cal A}_n$ so that
$$	\Phi_1*\Phi_2(z_1,\ldots,z_n)=\sum_{X\subset[n]}E_X^1E_X^2z^X      $$
With respect to $*$, ${\cal A}_n$ is thus isomorphic to the multiplicative semigroup of complex functions on $\{X:X\subset[n]\}$.  Note that $(\Phi_1*\Phi_2)^\dagger=\Phi_1^\dagger*\Phi_2^\dagger$.
\medskip
	We define the {\it inner radius} $r(\Phi)$, for $\Phi\in{\cal A}_n$, by $r(\Phi)=\infty$ if $\Phi$ is a constant $\ne0$, and by
$$	r(\Phi)
	=\sup\{r\ge0:\Phi(z_1\ldots,z_n)\ne0\quad{\rm if}\quad|z_1|,\ldots,|z_n|<r\}  $$
otherwise.  By compactness, it follows that $\Phi(\xi_1,\ldots,\xi_n)=0$ for some $\xi_1,\ldots,\xi_n$ with $|\xi_1|,\ldots,|\xi_n|\le r(\Phi)$.  If $|\xi_k|<r(\Phi)$ for  some $k$, we may relabel the $z_j$'s so that $\Phi(\xi_1,\ldots,\xi_n)$ $=0$ with $|\xi_1|,\ldots,|\xi_k|<r(\Phi)$, $|\xi_{k+1}|\ldots,|\xi_n|=r(\Phi)$ and $k<n$.  Then, if $k\ge1$,
$$	(z_1,\ldots,z_k)\mapsto\Phi(z_1,\ldots,z_k,\xi_{k+1},\ldots,\xi_n)      $$
must vanish identically.  Otherwise for any small change $(\xi_{k+1},\ldots,\xi_n)\to(\eta_{k+1}\ldots,\eta_n)$ we could find $(\eta_1,\ldots,\eta_k)$ close to $(\xi_1,\ldots,\xi_k)$ so that $\Phi(\eta_1,\ldots,\eta_n)=0$.  In particular we could take $|\eta_1|,\ldots,|\eta_n|$ $<r(\Phi)$, in contradiction with the definition of $r(\Phi)$.  A consequence of the above argument is that there are $\zeta_1,\ldots,\zeta_n$ with $\Phi(\zeta_1,\ldots,\zeta_n)=0$ and $|\zeta_1|,\ldots,|\zeta_n|=r(\Phi)$.
\medskip
	It will be convenient to denote by $[{\cal A}_n]_{\bf R},[{\cal A}_n]_{\bf C}$ the real and complex projective spaces associated with ${\cal A}_n$, {\it i.e.}, the set of nonzero classes $[\Phi]=\{\lambda\Phi:\lambda\ne0,\lambda\in{\bf R}$ or ${\bf C}\}$.  It is readily seen that the function $r(\cdot)$ is well defined on the classes $\in[{\cal A}_n]_{\bf C}$.
\medskip
	The set $LY_n$ mentioned in the introduction consists of those $\Psi\in{\cal A}_n$ such that $r(\Psi)\ge1$ and $r(\Psi^\dagger)\ge1$ (and therefore $r(\Psi)=r(\Psi^\dagger)=1$ by Proposition 2(b) below).
\medskip
	If $\Phi\in{\cal A}_n$, we define $\Psi_\Phi\in{\cal A}_{n+1}$ by
$$	\Psi_\Phi(z_1,\ldots,z_{n+1})=z_{n+1}\Phi^\dagger(z_1,\ldots,z_n)
	+\Phi(z_1,\ldots,z_n)      $$
It follows that $\Psi_\Phi^\dagger=\Psi_\Phi$ (where $\dagger$ denotes now a map ${\cal A}_{n+1}\to{\cal A}_{n+1}$), and $\Psi_{\Phi_1*\Phi_2}=\Psi_{\Phi_1}*\Psi_{\Phi_2}$ (where $*$ in the right-hand side is the product in ${\cal A}_{n+1}$).
\medskip\noindent
{\bf 2. Proposition} (properties of the inner radius $r(\Phi)$).
\medskip
	{\sl{\rm (a)} $[\Phi]\mapsto r([\Phi])$ is continuous $:[{\cal A}_n]_{\bf C}\to[0,\infty]$ (the compactification of the interval $[0,\infty)$).
\medskip\noindent
{\rm (b)} $r(\Phi^\dagger)\le r(\Phi)^{-1}$
\medskip\noindent
{\rm (c)} if $\Phi_1,\Phi_2\in{\cal A}_n$, then $r(\Phi_1*\Phi_2)\ge r(\Phi_1)r(\Phi_2)$
\medskip\noindent
{\rm (d)} $r(\Phi)>0$ if and only if $E_\emptyset\ne0$ (in particular, $r(\Phi)>0$ when $\Phi$ is $*$-invertible)
\medskip\noindent
{\rm (e)} if $r(\Phi)\ge1$, then $|E_X|\le|E_\emptyset|$ for all $X$
\medskip\noindent
{\rm (f)} for $\lambda\in{\bf C}\backslash\{0\}$ write $(\Phi\circ\lambda)(z_1,\ldots,z_n)=\Phi(\lambda z_1,\ldots,\lambda z_n)$, then $r(\Phi\circ\lambda)=|\lambda|^{-1}r(\Phi)$.}
\medskip
	(a) Let $\Phi\ne0$ and $\Phi(z_1\ldots,z_n)=0$ with $|z_1|,\ldots,|z_n|\le r(\Phi)$ finite.  Given $\epsilon>0$, if $\tilde\Phi$ is close to $\Phi$ we have $\tilde\Phi(\tilde z_1,\ldots,\tilde z_n)=0$ for some $(\tilde z_1,\ldots,\tilde z_n)$ close to $(z_1,\ldots,z_n)$, and therefore $|\tilde z_1|,\ldots,|\tilde z_n|\le r(\Phi)+\epsilon$, hence $r(\tilde\Phi)<r(\Phi)+\epsilon$.  Also, by compactness, we can find $a>0$ such that $|\Phi(z_1\ldots,z_n)|\ge2a$ if $|z_1|,\ldots,|z_n|\le r(\Phi)-\epsilon$.  Therefore, for $\tilde\Phi$ close to $\Phi$, we have $|\tilde\Phi(z_1\ldots,z_n)|\ge a$ if $|z_1|,\ldots,|z_n|\le r(\Phi)-\epsilon$, hence $r(\tilde\Phi)\ge r(\Phi)-\epsilon$.  Thus, finally, we have $|r(\tilde\Phi)-r(\Phi)|\le\epsilon$ for $\tilde\Phi$ close to $\Phi$ if $r(\Phi)$ is finite.  We have $r([\Phi])=\infty$ only when $[\Phi]$ is the class $[1]$ of nonzero constants, and it is readily seen that $r$ is also continuous at $[1]$.
\medskip
	(b) is trivial if $r(\Phi)$ or $r(\Phi^\dagger)$ vanishes.  In other cases $\Phi$ is non constant and $r(\Phi)>0$.  Then we have seen that there are $\zeta_1,\ldots,\zeta_n$ such that $\Phi(\zeta_1,\ldots,\zeta_n)=0$ with $|\zeta_1|,\ldots,|\zeta_n|=r(\Phi)>0$, and therefore $\Phi^\dagger(\zeta_1^{*-1},\ldots,\zeta_n^{*-1})=0$ with $|\zeta_1^{*-1}|,\ldots,|\zeta_n^{*-1}|=r(\Phi)^{-1}$, proving the assertion.
\medskip
	(c) is Corollary A3 of Appendix A.
\medskip
	(e) Let $E=\max_X|E_X|$ and write $\tilde\Phi=\Phi/E$.  Then $r(\tilde\Phi)\ge1$ and the coefficients $(E_X/E)^n$ of $\tilde\Phi^{*n}$ have a limit over a suitable subsequence $n\to\infty$.  Along this subsequence $\tilde\Phi^{*n}\to\Phi_0\in{\cal A}_n$, where $\Phi_0\ne0$.  We have $r(\tilde\Phi^{*n})\ge1$ by (c), thus $r(\Phi_0)\ge1$ by (a), and $\Phi_0(0,\ldots,0)\ne0$ by (d).  Since $\Phi_0(0,\ldots,0)=\lim(E_\emptyset/E)^n$ we cannot have $|E_\emptyset/E|<1$.  We have thus $|E_\emptyset|/E=1$, hence finally $|E_X|\le E=|E_\emptyset|$.
\medskip
	(d), and (f) are clear from the definitions.\qed
\medskip\noindent
{\bf 3. Theorem} (characterization of $LY_{n+1}$).
\medskip
	{\sl An element $\Psi\in{\cal A}_{n+1}$ is in $LY_{n+1}$ if and only if $\Psi=c\Psi_\Phi$ where $|c|=1$ and $\Phi\in{\cal A}_n$ satisfies $r(\Phi)\ge1$.}
\medskip
	{\bf First part of proof} ($\Psi\in LY_{n+1}\Rightarrow\Psi=c\Psi_\Phi$).
\medskip
	Let $\Psi\in LY_{n+1}$, ({\it i.e.}, $\Psi\in{\cal A}_{n+1}$, and $r(\Psi)\ge1,r(\Psi^\dagger)\ge1$).  We write ${\bf T}^n=\{(\alpha_1,\ldots,\alpha_n):|\alpha_1|=\ldots=|\alpha_n|=1\}$.  When $(\alpha_1,\ldots,\alpha_n)\in{\bf T}^n$, the functions
$$	z\quad\mapsto\quad\Psi(\alpha_1,\ldots,\alpha_n,z)\quad,\quad
	\Psi^\dagger(\alpha_1,\ldots,\alpha_n,z)      $$
are affine on ${\bf C}$.  If $z\mapsto\Psi(\alpha_1,\ldots,\alpha_n,z)$ is constant $\ne0$ (hence $\Psi(\alpha_1,\ldots,\alpha_n,1)\ne0$) then
$$	z\quad\mapsto\quad\Psi^\dagger(\alpha_1,\ldots,\alpha_n,z)
	=\alpha_1\cdots\alpha_nz[\Psi(\alpha_1,\ldots,\alpha_n,1)]^*      $$
is non constant, vanishing at $z=0$.  By a small change of $(\alpha_1,\ldots,\alpha_n,0)$ we obtain $(z_1,\ldots,z_n,z_{n+1})$ such that $\Psi^\dagger(z_1,\ldots,z_n,z_{n+1})=0$ and $|z_1|<1,\ldots,|z_{n+1}|<1$ in contradiction with $r(\Psi^\dagger)\ge1$.  Therefore, if $z\mapsto\Psi(\alpha_1,\ldots,\alpha_n,z)$ is constant, it is identically $0$.  Similarly, if $z\mapsto\Psi^\dagger(\alpha_1,\ldots,\alpha_n,z)$ is constant, it is identically $0$.
\medskip
	If the function $z\mapsto\Psi(\alpha_1,\ldots,\alpha_n,z)$ is not constant, it vanishes at $\beta\in{\bf C}$, and $z\mapsto\Psi^\dagger(\alpha_1,\ldots,\alpha_n,z)$ is non constant vanishing at $\beta^{*-1}$.  We cannot have $|\beta|<1$ because we could, by a small change of $(\alpha_1,\ldots,\alpha_n,\beta)$, obtain $(z_1,\ldots,z_n,z_{n+1})$ such that $\Psi(z_1,\ldots,z_n,z_{n+1})=0$ and $|z_1|<1,\ldots,|z_{n+1}|<1$ in contradiction with $r(\Psi)\ge1$. Similarly we cannot have $|\beta|>1$.
\medskip
	We have thus shown that $	z\mapsto\Psi(\alpha_1,\ldots,\alpha_n,z),\Psi^\dagger(\alpha_1,\ldots,\alpha_n,z)$ either are both identically $0$, or are both non constant and vanish at $z=\beta$ with $|\beta|=1$.  In the latter case, there is $C=C(\alpha_1,\ldots,\alpha_n)\ne0$ such that $\Psi^\dagger(\alpha_1,\ldots,\alpha_n,z)=C\Psi(\alpha_1,\ldots,\alpha_n,z)$.  Since $\Psi$ does not vanish identically and is continuous on ${\bf T}^{n+1}$, we can choose a nonempty open set ${\cal O}\in{\bf T}^{n+1}$ such that $\Psi,\Psi^\dagger$ do not vanish in ${\cal O}$, and therefore $\Psi^\dagger(\alpha_1,\ldots,\alpha_n,\alpha_{n+1})=C(\alpha_1,\ldots,\alpha_n)\Psi(\alpha_1,\ldots,\alpha_n,\alpha_{n+1})$ when $(\alpha_1,\ldots,\alpha_n,\alpha_{n+1})\in{\cal O}$.  We have thus $\Psi^\dagger=C\Psi$ in ${\cal O}$ with $C$ independent of the coordinate $\alpha_{n+1}$.  But similarly $C$ is independent of $\alpha_1,\ldots,\alpha_n$, and is therefore constant on ${\cal O}$.  By analyticity, this implies $\Psi^\dagger=C\Psi$ on ${\bf C}^{n+1}$.  
\medskip
	Note that $\Psi=\Psi^{\dagger\dagger}=(C\Psi)^\dagger=C^*\Psi^\dagger$ so that $C^{*-1}=C$ and $|C|=1$.  Choose $c$ such that $c^{-2}=C$, hence $|c|=1$.  Define $\Psi_c=c^{-1}\Psi$, then $\Psi_c^\dagger=c\Psi^\dagger=cc^{-2}\Psi=c^{-1}\Psi=\Psi_c$.  Writing
$$	\Psi_c(z_1,\ldots,z_{n+1})=\sum_{X\subset[n+1]}E_Xz^X
	=\sum_{X\subset[n]}E_Xz^X+z_{n+1}\sum_{X\subset[n]}F_Xz^X      $$
we see that $\Psi_c^\dagger=\Psi_c$ is equivalent to $F_X=E_{[n]\backslash X}^*$, or $\Psi_c=\Psi_\Phi$ with $\Phi(z_1,\ldots,z_n)=\sum_{X\subset[n]}E_Xz^X$.  We have thus shown that if $\Psi\in LY_{n+1}$, then $\Psi=c\Psi_\Phi$ with $|c|=1$ and $\Phi\in{\cal A}_n$.  Furthermore
$$	\Phi(z_1,\ldots,z_n)=c^{-1}\Psi(z_1,\ldots,z_n,0)      $$
and since tht right-hand side cannot vanish when $|z_1|,\ldots,|z_n|<1$, we have $r(\Phi)\ge1$.  This concludes the first part of the proof.
\medskip
	{\bf Second part of proof} ($\Psi\in LY_{n+1}\Leftarrow\Psi=c\Psi_\Phi$).
\medskip
	We will now show that $r(\Phi)\ge1$ implies that $\Psi_\Phi(z_1,\ldots,z_{n+1})\ne0$ if $|z_1|,\ldots,|z_{n+1}|<1$, {\it i.e.}, $r(\Psi_\Phi)\ge1$.  Since $\Psi_\Phi^\dagger=\Psi_\Phi$, this will imply that $c\Psi_\Phi\in LY_{n+1}$.
\medskip
	For $|z_1|,\ldots,|z_n|<1$, and $z_{n+1}=0$ we have $\Psi_\Phi(z_1,\ldots,z_n,0)=\Phi(z_1,\ldots,z_n)\ne0$.  If $z_{n+1}\ne0$, $\Psi_\Phi(z_1,\ldots,z_n,z_{n+1})=0$ is equivalent to
$$	z_{n+1}^{-1}=-{\Phi^\dagger(z_1,\ldots,z_n)\over\Phi(z_1,\ldots,z_n)}      $$
We thus have to prove that
$$	\Big|{\Phi^\dagger(z_1,\ldots,z_n)\over\Phi(z_1,\ldots,z_n)}\Big|\le1      $$
if $|z_1|,\ldots,|z_n|<1$.  Take $\lambda$ real $<1$, then $\Phi\circ\lambda\to\Phi,(\Phi\circ\lambda)^\dagger\to\Phi^\dagger$ pointwise when $\lambda\to1$.  Therefore
$$	\Big|{\Phi^\dagger(z_1,\ldots,z_n)\over\Phi(z_1,\ldots,z_n)}\Big|
	=\lim_{\lambda\to1}\Big|{(\Phi\circ\lambda)^\dagger(z_1,\ldots,z_n)
	\over(\Phi\circ\lambda)(z_1,\ldots,z_n)}\Big|      $$
where, using the maximum principle,
$$	\Big|{(\Phi\circ\lambda)^\dagger(z_1,\ldots,z_n)
	\over(\Phi\circ\lambda)(z_1,\ldots,z_n)}\Big|\le\max_{|\alpha_1|=\ldots=|	\alpha_n|=1}\Big|{(\Phi\circ\lambda)^\dagger(\alpha_1,\ldots,\alpha_n)
	\over(\Phi\circ\lambda)(\alpha_1,\ldots,\alpha_n)}\Big|=1      $$
This concludes the second part of the proof.\qed
\vfill\eject
\noindent
{\bf 4. Remarks.}
\medskip
	(a) When $P$ is a complex polynomial of order $n$, there is a unique $\Psi\in{\cal A}_n$ symmetric in its $n$ variables such that $P(z)=\Psi(z,\ldots,z)$.  If we write $P\in{\cal U}_n$ when all $n$ roots $\alpha_i$ of $P$ satisfy $|\alpha_i|=1$, then $P\in{\cal U}_n$ is equivalent to $\Psi\in LY_n$ by Grace's theorem A5.  Therefore Theorem 3 shows that:
\medskip
	{\sl $P\in{\cal U}_{n+1}$ if and only if $P(z)=c(zQ^\dagger(z)+Q(z))$, where $c\ne0$, $Q(z)=\sum_{\ell=0}^nC_\ell z^\ell$ has all its roots $\beta_i$ in $\{z:|z|\ge1\}\cup\infty$, and $Q^\dagger(z)=\sum_{\ell=0}^nC_{n-\ell}^*z^\ell$.}
\medskip
	(b) Let $\phi(s)=(az+b)/(cz+d)$ be a fractional linear transformation of the Riemann sphere (with $ad-bc\ne0$).  If $\Phi\in{\cal A}_n $ we may define $\Phi^\phi\in{\cal A}_n$ by the replacement $z_i\to(az_i+b)/(cz_i+d)$ in $\Phi(z_1,\ldots,z_n)$, and then chasing the denominators.  This yields new versions of Theorem A5 where the unit circle is replaced by the real or the imaginary axis, etc.  We shall not discuss here these new versions.
\medskip\noindent
{\bf 5. Proposition} (the set $[{\cal J}_n]$ of Lee-Yang classes and its interior).
\medskip
	{\sl Let ${\cal H}_n=\{\Phi\in{\cal A}_n:\Phi^\dagger=\Phi\}$ and $[{\cal H}_n]_{\bf R}$ be the corresponding real projective space, consisting of classes $[\Phi]=\{\lambda\Phi:\lambda\in{\bf R}\backslash\{0\}\}$ of elements $\Phi\in{\cal H}_n\backslash\{0\}$.  Also let ${\cal J}_n=LY_n\cap{\cal H}_n$ and $[{\cal J}_n]_{\bf R}=\{[\Phi]:\Phi\in{\cal J}_n\}$.  We have then
$$ {\cal H}_{n+1}=\{\Psi_\Phi:\Phi\in{\cal A}_n\}\eqno{(*)}      $$
$$ {\cal J}_{n+1}
	=\{\Psi_\Phi:\Phi\in{\cal A}_n\quad{\rm and}\quad r(\Phi)\ge1\}\eqno{(**)}    $$
Furthermore, the set $[{\cal J}_{n+1}]_{\bf R}$ is the closure in $[{\cal H}_{n+1}]_{\bf R}$ of its interior $\{[\Psi_\Phi]:r([\Phi])>1\}=[{\cal J}^\circ_{n+1}]_{\bf R}$ where ${\cal J}^\circ_{n+1}=\{\Psi\in{\cal H}_{n+1}:\Psi(z_1,\ldots,z_n,z_{n+1})\ne0\,{\rm when}\,|z_1|,\ldots,|z_n|\le1\,{\rm and}\,|z_{n+1}|<1\}$.}
\medskip
	We obtain $(*)$ directly from the definitions.
\medskip
	Given $c\Psi_\Phi\in{\cal A}_{n+1}$, we see that $c\Phi$ is uniquely determined by taking $z_{n+1}=0$, and (if $\Phi\ne0$) $c\Psi_\Phi\in{\cal H}_{n+1}$ if and only if $c$ is real.  Thus the elements of $LY_{n+1}\cap{\cal H}_{n+1}$ are of the form $\pm\Psi_\Phi=\Psi_{\pm\Phi}$ with $r(\pm\Phi)=r(\Phi)\ge1$, proving $(**)$.
\medskip
	The class $[\Psi_\Phi]$ is of the form $\{\Psi_{\tilde\Phi}:\tilde\Phi\in[\Phi]\}$ and $[\Phi]\mapsto[\Psi_\Phi]$ is thus a homeomorphism $[{\cal A}_n]_{\bf R}\to[{\cal H}_{n+1}]_{\bf R}$.  The set $[{\cal J}_{n+1}]_{\bf R}=\{[\Psi_\Phi]:r([\Phi])\ge1\}$ is homeomorphic to $\{[\Phi]:r([\Phi])\ge1\}$, which is closed (hence compact) by continuity of $r(\cdot)$.  Similarly, the set $\{[\Psi_\Phi]:r([\Phi])>1\}$ is homeomorphic to $\{[\Phi]:r([\Phi])>1\}$ which is open by continuity of $r(\cdot)$.
\medskip
	The map $\lambda\mapsto[\Phi\circ\lambda]$ is continuous near $\lambda=1$ so that, if $r([\Phi])=1$, we have $r([\Phi\circ\lambda])=|\lambda|^{-1}$ which may be $>1$ or $<1$ for $[\Phi\circ\lambda]$ close to $[\Phi]$.  This shows that $\{[\Psi_\Phi]:r([\Phi])>1\}$ is the interior of $[{\cal J}_{n+1}]_{\bf R}$, and $[{\cal J}_{n+1}]_{\bf R}$ is the closure of its interior. 
\medskip
	Finally we have to show that $\Psi_\Phi\in{\cal J}_{n+1}^\circ\Leftrightarrow r(\Phi)>1$.  Indeed $\Psi_\Phi\in{\cal J}_{n+1}^\circ\Rightarrow r(\Phi)>1$ because $\Phi(z_1,\ldots,z_n)=\Psi(z_1,\ldots,z_n,0)$.  Conversely, if $r(\Phi)>1$ and $\Psi_\Phi(z_1,\ldots,z_n,z)$ $=0$ with $|z_1|,\ldots,|z_n|\le1$, we must have $z\ne0$ and
$$	z^{-1}=-{\Phi^\dagger(z_1,\ldots,z_n)\over\Phi(z_1,\ldots,z_n)}      $$
where
$$	\Big|-{\Phi^\dagger(z_1,\ldots,z_n)\over\Phi(z_1,\ldots,z_n)}\Big|
	\le\max_{|\alpha_1|=\ldots=|\alpha_n|=1}
	\Big|{\Phi^\dagger(\alpha_1,\ldots,\alpha_n)
	\over\Phi(\alpha_1,\ldots,\alpha_n)}\Big|=1      $$
hence $|z|\ge1$, {\it i.e.}, $\Psi_\Phi\in{\cal J}_{n+1}^\circ$.\qed
\medskip\noindent
{\bf 6. Remarks.}
\medskip
	(a) Let $\Psi_1,\Psi_2\in LY_n$ (or ${\cal J}_n$, or ${\cal J}_n^\circ$), then $\Psi_1*\Psi_2\in LY_n$ (or ${\cal J}_n$, or ${\cal J}_n^\circ$) by Proposition 2(c).
\medskip
	(b) An element $\Psi$ of ${\cal H}_{n+1}$ may be written as
$$	\Psi(z_1,\ldots,z_{n+1})      $$
$$	=A(z_1,\ldots,z_{n-1})+B(z_1,\ldots,z_{n-1})z_n
	+C(z_1,\ldots,z_{n-1})z_{n+1}+D(z_1,\ldots,z_{n-1})z_nz_{n+1}      $$
where $A,B,C,D\in{\cal A}_{n-1}$ and $D=A^\dagger,C=B^\dagger$ (here $\dagger$ is defined in ${\cal A}_{n-1}$).  The condition $[\Psi]\in[{\cal J}_{n+1}^\circ]_{\bf R}$ is that $\Phi(z_1,\ldots,z_n)=A(z_1,\ldots,z_{n-1})+B(z_1,\ldots,z_{n-1})z_n$ does not vanish for $|z_1|,\ldots,|z_n|\le1$ or equivalently that $A(z_1,\ldots,z_{n-1})\ne0$ for $|z_1|,\ldots,|z_{n-1}|\le1$ and $|B(\alpha_1,\ldots,\alpha_{n-1})|<|A(\alpha_1,\ldots,\alpha_{n-1})|$ for $|\alpha_1|=\ldots=|\alpha_{n-1}|=1$.  Note that, in this condition, $B$ can equivalently be replaced by $C$, which corresponds to interchanging $z_n$ and $z_{n+1}$.  This shows that the asymmetric choice of the variable $z_{n+1}$ in our characterization (Theorem 3) of Lee-Yang polynomials is inessential.
\medskip
	(c) Let $E_X^\beta =\exp\beta W_X$ for $X\subset[n]$, with real $\beta>0$ and $W_X\in\{-\infty\}\cup{\bf C}$.  Write $\Phi^\beta(z_1,\ldots,z_n)=\sum_{X\subset[n]}E_X^\beta z^X$.  If $r(\Phi^1)>1$ there is an open interval $I\ni1$ such that $r(\Phi^\beta)>1$ for $\beta\in I$.  For integer $m$ sufficiently large we have $mI\supset(m-1,m+1)$ hence, by Proposition 2(c), $r(\Phi^\beta)>1$ for all $\beta\ge m$.  Thus, if $\Psi_{\Phi^1}\in{\cal J}_n^\circ$, then $\Psi_{\Phi^\beta}\in{\cal J}_n^\circ$ for all sufficiently large $\beta$.
\medskip
	(d) {\sl If $\Psi\in{\cal J}_{n+1}$ has nontrivial factorization: $\Psi(z_1,\ldots,z_{n+1})=\Psi'(z_1,\ldots,z_k)\times\Psi''(z_{k+1},\ldots,z_{n+1})$, then $\Psi\notin{\cal J}_{n+1}^\circ$.}
\medskip
	Indeed, we have $\Psi'\in LY_k,\Psi''\in LY_{n-k+1}$, and we may assume $\Psi'\in{\cal J}_k,\Psi''\in{\cal J}_{n-k+1}$.  Therefore we may choose $z_1,\ldots,z_{n+1}$ such that $\Psi'(z_1,\ldots,z_k)=0$ with $|z_1|,\ldots,|z_k|\le1$ and $|z_{k+1}|,\ldots,|z_{n+1}|<1$.
\medskip
	(e) ${\cal J}_{n+1}$ has full dimension in ${\cal H}_{n+1}$ (the dimension of ${\cal H}_{n+1}$ is the real dimension of ${\cal A}_n$, {\it i.e.}, $2^{n+1}$).  For $n>1$, this is strictly greater than the dimension $n(n+1)/2+n+1+1$ of the set of high-temperature polynomials of Theorem 9 below (with $b$ real), which is essentially the class of polynomials originally considered by Lee and Yang [3].
\medskip\noindent
{\bf 7. Examples.}
\medskip
	In what follows we study
$$	 \Psi(z_1,\ldots,z_{n+1})=\sum_{X\subset[n+1]}(\prod_UE_{UX})z^X      $$
for various choices or the $U$ and $E_{UX}$.
\medskip
	(a) {\sl For $U=\{j,k\}\subset[n+1]$, with $j<k$, write
$$	E_{UX}=\left\{\matrix{a_U\quad{\rm if}\quad j\in X,k\notin X\cr 
	a_U^*\quad{\rm if}\quad j\notin X, k\in X\cr 
	b_U\quad{\rm if}\quad j,k\notin X\cr 
	b_U^*\quad{\rm if}\quad j,k\in X\cr}\right.      $$
Then, if the complex $a_U,b_U$ satisfy $|a_U/b_U|\le1$ for all $U$, we have $\Psi\in{\cal J}_{n+1}$.}
\medskip
	Using Remark 6(a), it suffices to consider the case of a single $U$ such that $E_{UX}\ne1$, and we may thus take $n=1,U=[n+1]=\{1,2\}$.  Then $\Psi=\Psi_\Phi$ with $\Phi(z_1)=b_U+a_Uz_1$, hence $r(\Phi)\ge1$ by assumption, and $\Psi\in{\cal J}_{n+1}$ by Proposition 5.\qed
\medskip\noindent
[The standard situation [3] corresponds to taking $b_U=1$ and $a_U$ real.  The case of complex $a_U$ was considered by Beauzamy [2].  It is a consequence of Theorem 9 below that for polynomials of the class (a) we may write
$$	\Psi(z_1,\ldots,z_{n+1})
	=C\tilde\Psi(\alpha_1z_1,\ldots,\alpha_{n+1}z_{n+1})      $$
where $C,\alpha_1,\ldots,\alpha_{n+1}\in{\bf C}, |\alpha_1|=\ldots=|\alpha_{n+1}|=1$, and $\tilde\Psi$ is in the class (a) with real $a_U,b_U$ and $0\le a_U\le1,b_U=1$.]
\medskip
	(b) {\sl For $U=\{j,k,l\}\subset[n+1]$, write $E_{UX}=b_U$ if $U\cap X=\emptyset, =b_U^*$ if $U\subset X,=1$ otherwise.  Then, if $b_U$ is real $\ge1$ for all $U$, we have $\Psi\in{\cal J}_{n+1}$.}
\medskip
	It suffices to consider the case of a single $U$ such that $E_{UX}\ne1$, and we may thus take $n=2,U=[n+1]=\{1,2,3\}$.  Then $\Psi=\Psi_\Phi$ with
$$	\Phi(z_1,z_2)=b_U+z_1+z_2+z_1z_2=(1+z_1)(1+z_2)-(1-b_U)      $$
The set $\{(1-z_1)(1-z_2):|z_1|<1,|z_2|<1\}$ is bounded by the cardioid
$$	\Gamma=\{\rho e^{i\theta}:\rho=2(1+\cos\theta):\theta\in[-\pi,\pi]\}      $$
and does not contain $1-b_U$, which is real $\le0$ by assumption.  Therefore $r(\Phi)\ge1$.\qed
\medskip\noindent
[We note that the situation described in (b), with real $E_{UX}=E_{UX}^{(b)}$, is in fact contained in the situation described in (a), with  $E_{UX}=E_{UX}^{(a)}$.  Take indeed $b_{\{1,2\}}=b_{\{2,3\}}=b_{\{3,1\}}=b$ in (a), then if we take $b_{\{1,2,3\}}=b^2$ in (b) we have
$$	E_{\{1,2,3\}X}^{(b)}=b^{-1}E_{\{1,2,3\}X}^{(a)}      $$
because both sides are equal to $b^2$ if $|X|=0$ or $3$, and $1$ if $|X|=1$ or $2$.  Therefore $\Psi^{(b)}={\rm const.}\Psi^{(a)}$.  This equivalence fails if $b$ is not real].
\medskip
	(c) If we take $U=[n+1]=\{1,2,3\}$ in (b) and replace the real $b_U$ by $b(\beta)=e^{\beta W}$ with ${\rm Re}W>0$, ${\rm Im}W\ne0$, the set $\{b(\beta):\beta\hbox{ real}\}$ is a logarithmic spiral and $\{1-b(\beta):\beta>0\}$ intersects the region $(1-z_1)(1-z_2):|z_1|<1,|z_2|<1\}$ inside the cardioid $\Gamma$ in a sequence of intervals [the point $1-b(\beta)$ stays inside (resp. outside) of the cardioid for sufficiently small (resp. large) $\beta$].  Therefore, there are successive intervals of $\{\beta:\beta>0\}$ such that $\Psi\notin{\cal J}_3$ and $\Psi\in{\cal J}_3$.
\vfill\eject
	(d) {\sl For $U=\{j,k,l,m\}\subset[n+1]$, and $X\subset[n+1]$, write $E_{UX}=b_U$ if $U\cap X=\emptyset, =b_U^*$ if $U\subset X,=1$ otherwise.  Then, if $b_U$ is real $\ge2$ (or $=1$) for all $U$ we have $\Psi\in{\cal J}_{n+1}$.}
\medskip	
	It suffices to consider the case of a single $U$ such that $E_{UX}\ne1$, and we may take $n=3,U=[n+1]=\{1,2,3,4\}$.  Then $\Psi=\Psi_\Phi$ with
$$	\Phi(z_1,z_2,z_3)=(1+z_1)(1+z_2)(1+z_3)-(1-b_U)       $$
One checks that $\{(1-z_1)(1-z_2)(1+z_3):|z_1|<1,|z_2|<1,|z_3|<1\}$ intersects the real interval $(-\infty,0]$ in $(-1,0)$, so that for real $b_U\ge2$ (or $=1$) we have $r(\Phi)\ge1$.\qed
\medskip\noindent
{\bf 8. Lemma.}
\medskip
	{\sl For real $\beta>0$, and $W_X\in{\bf C}$ if $X\subset[n]$, write
$$	E_X^\beta=\exp\beta W_X\qquad,\qquad
	\Phi^\beta(z_1,\ldots,z_n)=\sum_{X\subset[n]}E_X^\beta z^X      $$
If $r(\Phi^\beta)\ge1$ for a sequence of $\beta$'s tending to $0$ from above, then
$$	W_X=\sum_{j\notin X}\sum_{k\notin X}W_{jk}
	+\sum_{j\notin X}W_j+W_0      $$
with suitable real $W_{jk}=W_{kj}\ge0,W_{jj}=0$, and complex $W_j,W_0$.}
\medskip
	For small $\beta$ we have $E_X^\beta=1+\beta W_X+O(\beta^2)$, hence
$$	\Phi^\beta(z_1,\ldots,z_n)
	=\prod_1^n(1+z_j)+\beta\sum_XW_Xz^X+O(\beta^2)\sum_X|z|^X      $$
Writing $\zeta_j=1+z_j$ gives (assuming all $\zeta_j\ne0$, and $\sum|\zeta_j|$ bounded)
$$	\Phi^\beta(z_1,\ldots,z_n)
=\prod_1^n\zeta_j+\beta\sum_XW'_X\zeta^X+O(\beta^2)\sum_X|\zeta|^X      $$
$$	=(\prod_1^n\zeta_j)\Big[1+\beta\sum_XW''_X(\zeta^{-1})^X
	+O(\beta^2)\sum_X|\zeta^{-1}|^X\Big]      $$
where the $W'_X,W''_X$ are linear combinations of the $W_X$ with coefficients in ${\bf Z}$.
\medskip
	Taking $\gamma=\beta^{1/n},\zeta_1=\ldots=\zeta_n=\zeta=\gamma\theta^{-1}$, and writing $|X|=\hbox{card}\,X$, we have
$$	\Phi^\beta(z_1,\ldots,z_n)=\zeta^n\Big[1+W''_{[n]}\theta^n
	+\sum_{|X|<n}W''_X\gamma^{n-|X|}\theta^{|X|}
	+O(\gamma^{2n})\sum_X\gamma^{-|X|}\theta^{|X|}\Big]      $$
$$	=\zeta^n\Big[1+W''_{[n]}\theta^n+O(\gamma)\Big]      $$
for bounded $\theta$.  If $W_{[n]}\ne0$ there are thus zeros of $\Phi^\beta$ of the form $z_1=\ldots=z_n=\zeta-1$ with
$$	\zeta\approx(-W''_{[n]})^{1/n}\gamma\times n\hbox{-th root of }1      $$
[use the implicit function theorem].  For small $\gamma$ and $n>2$ this implies that $\Phi^\beta(z_1,\ldots,z_n)$ $=0$ for some $z_1,\ldots,z_n$ with $|z_1|,\ldots,|z_n|<1$.  If $r(\Phi^\beta)\ge1$ and $n>2$ we have thus $W''_{[n]}=0$.
\medskip
	Put the variables $z_j$, with $j\notin X$, equal to $0$ in $\Phi^\beta$, and consider the function of the remaining variables $z_k$, with $k\in X$.  Taking now $\gamma=\beta^{1/|X|}$ we see as above that $\Phi^\beta$ vanishes when $z_j=0\hbox{ if }j\notin X, z_k=-1+\zeta\hbox{ if }k\in X$, where
$$	\zeta\approx(-W''_X)^{1/|X|}\gamma\times|X|\hbox{-th root of }1      $$
Assuming $r(\Phi^\beta)\ge1$ we have thus $W''_X=0$ if $|X|>2$, and $W''_X$ real $\ge0$ if $|X|=2$ (so that $(-W''_X)^{1/2}$ is pure imaginary).
\medskip
	If $r(\Phi^\beta)\ge1$ for some arbitrarily small $\beta>0$, we have thus
$$	\sum W_Xz^X=(\prod_1^n\zeta_j)\sum W''_X(\zeta^{-1})^X
	=(\prod_1^n\zeta_j)\sum_{X:|X|\le2} W''_X(\zeta^{-1})^X      $$
$$	=\sum_{X:|X|\le2} W''_X\zeta^{[n]\backslash X}
	=\sum_{X:|X|\le2} W''_X(1+z)^{[n]\backslash X}      $$
$$	=\sum_Yz^Y\Big[\sum_{\{j,k\}\subset [n]\backslash Y}W''_{\{j,k\}}+\sum_{\{j\}\subset [n]\backslash Y}W''_{\{j\}}+W''_\emptyset\Big]      $$
and the lemma follows, with $W_{jk}={1\over2}W''_{\{j,k\}},W_j=W''_{\{j\}}$ and $W_0=W''_\emptyset$.\qed
\medskip\noindent
{\bf 9. Theorem} (high-temperature Lee-Yang polynomials).
\medskip
	{\sl Let $W_X\in{\bf C},E_X^\beta=\exp\beta W_X$ and $\Psi^\beta(z_1,\ldots,z_n)=\sum_{X\subset[n]}E_X^\beta z^X$.  We say that $(\Psi^\beta)_{\beta>0}$ is a high-temperature Lee-Yang polynomial if
\medskip\noindent
{\rm (a)} $\Psi^\beta\in LY_n$ for some sequence of real $\beta$'s tending to zero from above.
\medskip
	We claim that {\rm (a)} is equivalent to {\rm (b)} and also to {\rm (c)}:
 \medskip\noindent
{\rm (b)}	$\Psi^\beta\in LY_n$ for all $\beta>0$
\medskip\noindent
{\rm (c)}	there exist $W_{jk}\in{\bf R}, a_j\in{\bf R},b\in{\bf C}$ such that $W_{jk}=W_{kj}\ge0$ and
$$	W_X=-\sum_{j\in X}\sum_{k\notin X}W_{jk}-i\sum_{j\in X}a_j+b      $$}
\medskip
	The implication (c)$\Rightarrow$(b) is proved in Section {\bf 7}(a), and (b)$\Rightarrow$(a) is obvious.
\medskip
	We assume now that (a) holds.  Since $\Psi^\beta$ satisfies the conditions of Lemma 8 we have
$$	W_X=\sum_{j\notin X}\sum_{k\notin X}W_{jk}+\sum_{j\notin X}W_j+W_0
	=\sum_{j\notin X}\sum_kW_{jk}-\sum_{j\notin X}\sum_{k\in X}W_{jk}
	+\sum_{j\notin X}W_j+W_0      $$
$$	=-\sum_{j\in X}\sum_{k\notin X}W_{jk}
	+\sum_{j\notin X}\tilde W_j+W_0      $$
with real $W_{jk}=W_{kj}\ge0,\tilde W_j=W_j+\sum_kW_{jk}\in{\bf C},W_0\in{\bf C}$.  In view or Theorem 3 the quantity $E_X^\beta/(E_{[n]\backslash X}^\beta)^*$ is independent of $X$, hence the limit for $\beta\to0$ of 
$$	{d\over d\beta}\exp\beta(W_X-W_{[n]\backslash X}^*)
	={d\over d\beta}\exp\beta
	(\sum_{j\notin X}\tilde W_j-\sum_{j\in X}\tilde W_j^*+W_0-W_0^*) $$
$$	=(\sum_{j\notin X}\tilde W_j-\sum_{j\in X}\tilde W_j^*+W_0-W_0^*)
	\exp\beta(\sum_{j\notin X}\tilde W_j-\sum_{j\in X}\tilde W_j^*+W_0-W_0^*)      $$
is also independent of $X$.  Since this limit is $\sum_{j\notin X}\tilde W_j-\sum_{j\in X}\tilde W_j^*+W_0-W_0^*$, we have $\tilde W_j+\tilde W_j^*=0$, {\it i.e.}, $\tilde W_j=ia_j$ with $a_j\in{\bf R}$.  Writing $W_0+i\sum_ja_j=b\in{\bf C}$, we have thus proved (a)$\Rightarrow$(c).\qed
\medskip\noindent
{\bf 10. Higher degree.} 
\medskip
	Given integers $m_1,\ldots,m_n\ge1$, let ${\bf m}=(m_1,\ldots,m_n)$ and ${\cal B}_{\bf m}\subset{\bf C}[z_1,\ldots,z_n]$ consist of polynomials separately of degree $m_1,\ldots,m_n$ in $z_1,\ldots,z_n$.  A polynomial $\Phi\in{\cal B}_{\bf m}$ is thus of the form
$$	\Phi(z_1,\ldots,z_n)=\sum_{k_1=0}^{m_1}\ldots\sum_{k_n=0}^{m_n}
	E_{k_1\cdots k_n}{m_1!\over k_1!(m_1-k_1)!}z_1^{k_1}\cdots
	{m_n!\over k_n!(m_n-k_n)!}z_n^{k_n}      $$
We define $\Phi^\dagger\in{\cal B}_{\bf m}$ by
$$	\Phi^\dagger(z_1,\ldots,z_n)=\sum_{k_1=0}^{m_1}\ldots\sum_{k_n=0}^{m_n}
	E_{k_1\cdots k_n}^*{m_1!\over k_1!(m_1-k_1)!}z_1^{m_1-k_1}\cdots
	{m_n!\over k_n!(m_n-k_n)!}z_n^{m_n-k_n}      $$
We also let $r(\Phi)$ be $\infty$ if $\Phi$ is a constant $\ne0$, and
$$	r(\Phi)
	=\sup\{r\ge0:\Phi(z_1\ldots,z_n)\ne0\quad{\rm if}\quad|z_1|,\ldots,|z_n|<r\}  $$
otherwise.  With this notation we may define a set $LY_{\bf m}$ of Lee-Yang polynomials $\Psi$ in ${\cal B}_{\bf m}$ by $r(\Psi)=r(\Psi^\dagger)=1$
\medskip
	Let now ${\cal A}_{\bf m}\subset{\bf C}[z_{11},\ldots,z_{1m_1},\ldots,z_{n1},\ldots,z_{nm_n}]$ consist of the multiaffine polynomials invariant under permutations of $z_{i1},\ldots,z_{im_i}$ for $i=1,\ldots,n$.  There is a unique linear isomorphism $\sigma:{\cal B}_{\bf m}\to{\cal A}_{\bf m}$ such that $(\sigma^{-1}\Psi)(z_1\ldots,z_n)$ is obtained if one replaces $z_{i1},\ldots,z_{im_i}$ by $z_i$ in $\Psi$.  We note that $\sigma\Phi^\dagger=(\sigma\Phi)^\dagger$, and (using Grace's theorem A5) that $r(\Phi)=r(\sigma\Phi)$.  If $\Phi_1,\Phi_2\in{\cal B}_{\bf m}$ we also define $\Phi_1*\Phi_2$ by $\sigma(\Phi_1*\Phi_2)=\sigma\Phi_1*\sigma\Phi_2$, and we have $r(\phi_1*\Phi_2)\ge r(\Phi_1)r(\Phi_2)$.
\vfill\eject
	With the above notation we see that questions about $LY_{\bf m}$ are transformed by use of the isomorphism $\sigma$ into questions about $LY_{|{\bf m}|}$ where $|{\bf m}|=m_1+\ldots+m_n$.  For instance, one finds that if $\Psi\in LY_{\bf m}$, then $\Psi=c\tilde\Psi$ where $|c|=1$ and $\tilde\Psi\in LY_{\bf m},\tilde\Psi^\dagger=\tilde\Psi$.
\medskip\noindent
{\bf 11. Physical interpretation.}
\medskip
	The polynomials $\Psi(z_1,\ldots,z_{n+1})=\sum_{X\subset[n+1]}E_Xz^X$ relevant to physics are such that $E_X=e^{\beta W_X}$, where $\beta,W_X\in{\bf R}$, and $\beta^{-1}>0$ is interpreted as {\it temperature}, while $-W_X$ is the {\it energy} of the {\it configuration} $X$.  For such $\Psi$, we have $\Psi\in LY_{n+1}$ if and only if $\Psi=\Psi_\Phi$ and $r(\Phi)\ge1$, {\it i.e.}, $\Psi\in{\cal J}_{n+1}$ (Theorem 3 and Proposition 5).  Writing $\Psi=\Psi^\beta$, we find (Remark 6(c)) that if $\Psi^\beta\in{\cal J}_{n+1}^\circ$ (the interior of ${\cal J}_{n+1}$) for some $\beta$, then $\Psi^\beta\in{\cal J}_{n+1}^\circ$ for all sufficiently large $\beta$ ({\it i.e.}, small temperatures).  There are cases where $\Psi^\beta\in{\cal J}_{n+1}^\circ$ for small temperatures and $\Psi^\beta\notin{\cal J}_{n+1}$ for large temperatures (Example 7(d)).  [For complex $W_X$ we may have several successive intervals of temperature where $\Psi^\beta$ is in and out of ${\cal J}_{n+1}$ (Example 7(c)), I do not know if this can happen for real $W_X$].  If $\Psi^\beta\in{\cal J}_{n+1}$ for a sequence of $\beta$'s tending to 0 (high temperatures), then $\Psi^\beta\in{\cal J}_{n+1}$ for all $\beta>0$ (all temperatures) and
$$	W_X=-\sum_{j\in X}\sum_{k\notin X}W_{jk}+b      $$
with $W_{jk},b\in{\bf R}$ such that $W_{jk}=W_{kj}\ge0$ (Theorem 9).  [Since $\Psi^\beta(0,\ldots,0)=e^{\beta W_\emptyset}=e^{\beta b}$, we have $b\in{\bf R}$.  Taking only $z_j\ne0$ in $\Psi^\beta$ yields $e^{-i\beta a_j}z_j$, hence $a_j=0$].  This means that those $\Psi^\beta$ which are Lee-Yang polynomials at high temperatures, hence at all temperatures, are precisely of the form considered by Lee and Yang [3].
\medskip
	The higher degree situation of Section 10 corresponds physically to {\it higher spins} (the degree $m_i$ corresponds to spin $(m_i+1)/2$).  The $\Psi^\beta\in{\cal B}_{\bf m}$ relevant to physics which are Lee-Yang polynomials at high temperature, hence at all temperatures, can again be determined.  Using the notation of Section 10, they correspond to $E_{k_1\ldots k_n}=\exp\beta W_{k_1\ldots k_n}$ (with $\beta>0$, $W_{k_1\ldots k_n}\in{\bf R}$) such that
$$	W_{k_1\ldots k_n}=-\sum_{i=1}^n\sum_{j=1}^nW_{ij}k_i(m_j-k_j)+b      $$
with $W_{ij},b\in{\bf R},W_{ij}=W_{ji}\ge0$ ($W_{ii}$ in general does not vanish).
\medskip\noindent
{\bf A. Appendix} (some properties of multiaffine polynomials).\footnote{*}{See [6] and references quoted there.} 
\medskip\noindent
{\bf A1. Lemma} [5].
\medskip
	{\sl Let $K_1,K_2$ be closed subsets of ${\bf C}$, with $K_1,K_2\not\ni0$.  If $\Phi\in{\cal A}_2$ and
$$	\Phi(z_1,z_2)\equiv A+Bz_1+Cz_2+Dz_1z_2\ne0      $$
whenever $z_1\notin K_1$ and $z_2\notin K_2$.  Then
$$	\tilde\Phi(z)\equiv A+Dz\ne0      $$
whenever $z\notin-K_1\cdot K_2$.  {\rm [}We have written $-K_1\cdot K_2=\{-uv:u\in K_1,v\in K_2\}${\rm ]}.}
\medskip
	The map $\Phi\mapsto\tilde\Phi$ of ${\cal A}_2$ to ${\cal A}_1$ is called {\it Asano contraction}.  For the convenience of the reader we reproduce here the easy proof of this lemma (which generalizes an earlier result of Asano [1]).
\medskip
	Since $K_1,K_2\not\ni0$, we have $A\ne0$.  If $D=0$, there is nothing to prove.  If $D\ne0$ and $AD-BC=0$, we have
$$	 A+Bz_1+Cz_2+Dz_1z_2=D(z_1+{C\over D})(z_2+{A\over C})      $$
which implies $-C/D\in K_1$, $-A/C\in K_2$, hence $A/D\in K_1\cdot K_2$ and $A+Dz=0$ only if $z=-A/D\in -K_1\cdot K_2$.
\medskip
	Suppose now that $D\ne0$ and $AD-BC\ne0$, and write\footnote{*}{This form of the argument was communicated to me by F.J. Dyson.}
$$	\phi(z)=-{A+Bz\over C+Dz}\qquad,\qquad\psi(z)={A\over Dz}      $$
where $\phi$, $\psi$ are considered as mappings of the Riemann sphere (add a point at infinity to ${\bf C},K_1,K_2$).  If we write $\omega=\phi\psi^{-1}$, $z_2=\omega z_1$ is equivalent to
$$	AB+ADz_1+ADz_2+CDz_1z_2=0      $$
showing that $\omega$ is an involution.  Since $K_2$ is a proper closed set, $\omega K_2$ cannot be interior to $K_2$ [otherwise $\omega^2 K_2$ would be interior to $K_2$].  Thus
$$	(\omega K_2)\cap({\bf C}\backslash K_2)^{-}\ne\emptyset      $$
where $({\bf C}\backslash K_2)^{-}$ denotes the closure of the complement of $K_2$.  By assumption ${\bf C}\backslash K_2\subset\phi K_1$ and, since $\phi K_1$ is closed, $({\bf C}\backslash K_2)^{-}\subset\phi K_1$, hence $\omega K_2\cap\phi K_1\ne\emptyset$, or $\phi\psi^{-1}K_2\cap\phi K_1\ne\emptyset$, or $K_2\cap\psi K_1\ne\emptyset$, {\it i.e.},
$$	(\exists z)\Big({A\over Dz}\in K_1\hbox{ and }z\in K_2\Big)      $$
so that $A/D\in K_1\cdot K_2$ and finally $-A/D\in -K_1\cdot K_2$.\qed
\medskip\noindent
{\bf A2. Proposition} [5].
\medskip
	{\sl Let $K_{ij}$ be a closed subset of ${\bf C}$ with $K_{ij}\not\ni0$, for $i=1,2$ and $j=1,\ldots,n$.  If $\Phi_1,\Phi_2\in{\cal A}_n$, and $\Phi_i(z_1,\ldots,z_n)\ne0$ whenever $z_1\notin K_{i1},\ldots,z_n\notin K_{in}$, then $\Phi_1*\Phi_2(z_1,\ldots,z_n)\ne0$ whenever $z_1\notin-K_{11}\cdot K_{21},\ldots,z_n\notin-K_{1n}\cdot K_{2n}$.}
\medskip
	To show this, start from the element of ${\cal A}_{2n}$ defined by the product $\Phi_1(z_{11},\ldots,z_{1n})$ $\times\Phi_2(z_{21},\ldots,z_{2n})$, different from 0 whenever $z_{ij}\notin K_{ij}$ for $i=1,2;j=1,\ldots,n$.  Then perform successive Asano contractions $(z_{1k},z_{2k})\to z_k$ for $k=1,\ldots,n$.  At the $k$-th step we have an element of ${\cal A}_{2n-k}$.  Induction on $k$ and use of Lemma A1 yield that this element of ${\cal A}_{2n-k}$ is $\ne0$ when $z_1\notin-K_{11}\cdot K_{21},\ldots,z_k\notin-K_{1k}\cdot K_{2k},z_{1,k+1}\notin K_{1,k+1},z_{2,k+1}\notin K_{2,k+1},\ldots,z_{1n}\notin K_{1n},z_{2n}\notin K_{2n}$.  Taking $k=n$ proves the proposition.\qed
\medskip\noindent
{\bf A3. Corollary}.
\medskip
	{\sl If $\Phi_1,\Phi_2\in{\cal A}_n$, then $r(\Phi_1*\Phi_2)\ge r(\Phi_1)r(\Phi_2)$.}
\medskip
	This follows by taking $K_{ij}=\{z:|z|\ge r(\Phi_i)\}$.\qed
\medskip\noindent
{\bf A4. Remark}.
\medskip
	Asano [1] originally considered the situation where $K_1,K_2=\{z:|z|\ge1\}$ in Lemma 1.  In that case, the fact that
$$	 A+Bz_1+Cz_2+Dz_1z_2\ne0      $$
if $z_1\notin K_1,z_2\notin K_2$, implies that the roots of $A+(B+C)z+Dz^2$ have absolute value $\ge1$, and the same is true of their product $A/D$, proving the lemma.
\medskip\noindent
{\bf A5. Theorem} (Grace's theorem).
\medskip
	{\sl Let $P$ be a complex polynomial of degree $n$ in one variable and $\Phi\in{\cal A}_n$ be the only polynomial symmetric in its $n$ arguments such that
$$	\Phi(z,\ldots,z)=P(z)      $$
If the $n$ roots of $P$ are contained in a closed circular region $K$ and $z_1\notin K,\ldots,z_k\notin K$, then $\Phi(z_1,\ldots,z_n)\ne0$.}
\medskip
	A closed circular region is a closed subset $K$ of ${\bf C}$ bounded by a circle or a straight line.  We allow the coefficients of $z^n,z^{n-1},\ldots$ in $P$ to vanish: we then say that some  of the roots of $P$ are at $\infty$, and we take $K$ noncompact.
\medskip
	For a proof see Polya and Szeg\"o [4] V, Exercise 145.\qed
\medskip\noindent
{\bf References.}
\medskip
	[1] T. Asano.  "Theorems on the partition functions of the Heisenberg ferromagnets.'' J. Phys. Soc. Jap. {\bf 29},350-359(1970).
\medskip
	[2] B. Beauzamy.  "On complex Lee and Yang polynomials." Commun. Math. Phys. {\bf 182},177-184(1996).
\medskip
	[3] T. D. Lee and C. N. Yang.  "Statistical theory of equations of state and phase relations. II.  Lattice gas and Ising model.''  Phys. Rev. {\bf 87},410-419(1952).
\medskip
	[4] G. Polya and G. Szeg\"o.  {\it Problems and theorems in analysis II.}  Springer, Berlin, 1976.
\medskip
	[5] D. Ruelle.  "Extension of the Lee-Yang circle theorem.''  Phys. Rev.
Letters {\bf 26},303-304(1971).
\medskip
	[6] D. Ruelle.   "Grace-like polynomials.'' in {\it Foundations of computational mathematics.  Proceedings of Smalefest 2000} edited by F. Cucker and J.M. Rojas, World Scientific, Singapore, 2002.
\end